%% file: custom.tex
\pdfoutput=1

\documentclass[11pt]{article}

\usepackage[final]{acl}
\usepackage{bm}
\usepackage{times}
\usepackage{latexsym}
\usepackage{amsmath}
\usepackage{amsfonts}
\usepackage{booktabs}
\usepackage{makecell}
\usepackage{multirow}
\usepackage[capitalise]{cleveref}
\crefname{section}{Sec.}{Secs.}

\usepackage[T1]{fontenc}

\usepackage[utf8]{inputenc}

\usepackage{microtype}

\usepackage{inconsolata}

\usepackage{graphicx}

\title{DNCASR: End-to-End Training for Speaker-Attributed ASR}

\author{\textbf{Xianrui Zheng}\textsuperscript{1}, \textbf{Chao Zhang}\textsuperscript{2}, \textbf{Philip C. Woodland}\textsuperscript{1}
\\
\\
 \textsuperscript{1}Department of Engineering, University of Cambridge, Trumpington St., Cambridge, UK.,
 \\
 \textsuperscript{2}Department of Electronic Engineering, Tsinghua University, Beijing, China
\\
 \texttt{
 \textsuperscript{1}{\{xz396, pw117\}@cam.ac.uk},  \textsuperscript{2}{cz277@tsinghua.edu.cn}
 }
}

\begin{document}
\maketitle
\input{sections/abstract}
\input{sections/intro}
\input{sections/related_work}

\input{sections/method.tex}
\input{sections/exp_setup}
\input{sections/results}

\input{sections/conclusion}

\section{Limitations}
DNCASR jointly fine-tunes a speaker clustering module (DNC) and an ASR module, but it still relies on a separate speaker embedding extraction module and a separate VAD. 
Our current setup has not investigated the possibility of jointly training the DNCASR with the speaker embedding extraction module and VAD.
For experiments on the real meeting corpus AMI, we use the AMI-MDM training data as the sole source of supervised training for the entire system. This makes it challenging to compare our system with others that rely on significantly more supervised data, much of which is not publicly available.  We have not yet tested the method on other multi-taker datasets.
Our experiments show that using an oracle ASR word tokens can significantly improve results. 
However, only a relatively small pre-trained WavLM model was used as the Wav encoder in the main experiments. 
With a larger Wav encoder pre-trained on more data, the ASR performance could be further enhanced, which would lead to improved speaker-attributed ASR results as shown in \cref{app:wavlm-large} even with the same amount of supervised training data.

\section{Ethics}

In general, an improved long form multi-speaker ASR in general could lead to decreased privacy in speech based communications (electronic and face-to-face), unless certain legal frameworks are developed and enforced.
However, our model uses speaker data from publicly available sources, focusing solely on generating speaker indices within meetings. 
It does not attempt to associate any data with specific speaker identities. 
This approach ensures privacy by preventing the identification of individual speakers. 
All data is anonymised, maintaining compliance with ethical standards while utilising publicly accessible data for model training and evaluation.
A positive impact would be better efficiency, transparency and traceability of speech based communications via easier access to information in consensually made recordings. 

\section{Acknowledgments}

Xianrui Zheng is supported by an Amazon Studentship. This work has been performed using resources provided by the Cambridge Tier-2 system operated by the University of Cambridge Research Computing Service (www.hpc.cam.ac.uk) funded by EPSRC Tier2 capital grant EP/T022159/1.

\bibliography{custom}
\appendix
\input{sections/appendix}

\end{document}

%% file: sections/abstract.tex
\begin{abstract}
This paper introduces DNCASR, a novel end-to-end trainable system designed for joint neural speaker clustering and automatic speech recognition (ASR), enabling speaker-attributed transcription of long multi-party meetings. 
DNCASR uses two separate encoders to independently encode global speaker characteristics and local waveform information, along with two linked decoders to generate speaker-attributed transcriptions.
The use of linked decoders allows the entire system to be jointly trained under a unified loss function.
By employing a serialised training approach, DNCASR effectively addresses overlapping speech in real-world meetings, where the link improves the prediction of speaker indices in overlapping segments.
Experiments on the AMI-MDM meeting corpus demonstrate that the jointly trained DNCASR outperforms a parallel system that does not have links between the speaker and ASR decoders. 
Using cpWER to measure the speaker-attributed word error rate, DNCASR achieves a 9.0\% relative reduction on the AMI-MDM Eval set.
\end{abstract}

%% file: sections/intro.tex
\section{Introduction}
\label{sec:intro}

To transcribe multi-speaker conversations, such as meetings and social gatherings, it is essential to not only recognise the spoken words but also attribute them to the correct speakers.
Systems for the task ``who spoke what'' often include two sub-systems: one is a diarisation sub-system \cite{tranterOverviewAutomatic2006,sellDiarizationHard2018,parkReviewSpeaker2021}, the other is an automatic speech recognition (ASR) sub-system \cite{watanabeHybridCTC2017,prabhavalkarEndtoendSpeech2024}. 
The diarisation task aims to identify ``who spoke when'', involving finding speaker indices, which are unique identifiers of speakers within each meeting e.g. 0, 1, rather than absolute speaker identities. 
Traditional speaker diarisation systems often consist of three main components: voice activity detection (VAD) \cite{sohnStatisticalModelbased1999,wangImprovedDNNbased2016}, speaker embedding extraction \cite{dehakFrontendFactor2011,snyderXvectorsRobust2018,dawalatabadECAPATDNNEmbeddings2021,koluguriTitanetNeural2022}, and speaker clustering \cite{chenSpeakerEnvironment1998,ningSpectralClustering2006}. 
The VAD module removes non-speech regions from the audio, leaving only the speech regions, referred to as VAD segments. 
The embedding extraction module generates speaker embeddings from VAD segments, and the clustering module assigns speaker indices to each embedding, forming speaker-homogeneous segments.
In cascaded speaker-attributed ASR systems \cite{rajIntegrationSpeech2021,zhengTandemMultitask2022,cornell2023chime}, speaker-homogeneous segments from the diarisation sub-system are sent to an ASR sub-system to decode, producing speaker-attributed transcriptions.
In cascaded systems, diarisation and ASR sub-systems are trained independently with no consideration of their interactions.

Several studies have investigated using neural network-based integrated systems for multi-talker ASR, moving away from cascaded approaches. 
One approach involves employing multiple output heads in an ASR model, where each head outputs words spoken by different speakers from a multi-speaker speech segment.
These models can be trained using a single ASR loss function 
\cite{changEndtoendMultispeaker2020,luStreamingEndtoend2021,sklyarStreamingMultispeaker2021,sklyarMultiturnRNNt2022},
or with an additional loss to guide the model to separate clean speech before recognising them \cite{sekiPurelyEndtoend2018,rajSURT202023}.
Although these systems can separately decode speech from multiple speakers within a segment, even when there is overlapping speech, they cannot assign speaker indices across an entire meeting.
Within a segment, there's also no guarantee that multiple turns from the same speaker will be output from the same output head. 
One possible method to extend multi-talker ASR to speaker-attributed ASR is to use a speaker inventory \cite{kandaJointSpeaker2020,luStreamingMultitalker2021}.
However, this approach requires pre-existing speaker profiles to be known in advance. 

For speaker-attributed ASR without a speaker inventory, a key challenge arises with long meetings: neural network-based systems often struggle to process the entire meeting at once, especially when the input is the full meeting waveform.
However, in order to assign speaker indices, the system must consider the entire meeting.
To address this issue, \citet{kandaTranscribediarizeNeural2022} proposed to jointly train ASR with speaker embeddings. 
The ASR module uses serialised output training (SOT) \cite{kandaSerializedOutput2020}, which generates transcriptions for all speakers within a VAD segment sequentially.
For each speaker turn, a jointly trained speaker decoder decodes a speaker embedding, then embeddings of all speaker turns across an entire meeting are clustered using a non-neural clustering algorithm to assign speaker indices. 

To enable end-to-end trainable speaker-attributed ASR, a neural clustering module is essential for generating speaker indices across the entire meeting. 
Several studies have proposed end-to-end neural diarisation systems that take full conversation waveforms or high-resolution filterbank features as input and directly output speaker indices \cite{Fujita2019Interspeech,landiniDiaPerEndEnd2024,horiguchiEncoderdecoderBased2022}.
EEND \cite{Fujita2019Interspeech} is a system that can perform diarisation in an end-to-end manner, integrating VAD, speaker embedding extraction, and clustering into a single model.
EEND takes an input of the high-resolution filterbank features of an entire conversation, and outputs a sequence of frame-level speaker activity probabilities.
\citet{cornellOneModel2024} combines EEND with an SOT ASR to perform speaker-attributed transcription.
However, EEND cannot process the waveform of the entire conversation when the duration is long. 
For long meetings such as those in the AMI dataset \cite{carlettaAmiMeeting2006}, the EEND and the speaker-attributed ASR system based on EEND have to split the meeting into short segments, and use a non-neural clustering algorithm to assign speaker indices across segments \cite{eend-vector-clustering}.
To handle entire meetings and generate speaker indices, some neural network-based diarisation systems use speaker embeddings as input instead of waveforms \cite{zhangFullySupervised2019,liDiscriminativeNeural2020,zhengSOTTriggered2024}. 
Each embedding either represents an entire utterance or a fixed-size window spanning several seconds of waveform. 
This method is more scalable for long meetings than processing the full waveform. 
However, speaker embeddings cannot directly produce speaker-attributed transcriptions because they discard word-level information, which requires high time-resolution input.

In this paper, we propose an end-to-end trainable speaker-attributed ASR approach, which produces words with speaker indices for the entire meeting, without relying on non-neural clustering algorithms.
The system, referred to as DNCASR, jointly trains a neural clustering module for generating speaker indices, and an ASR module for transcribing the spoken words.
During training, both the clustering and ASR modules can be optimised using a single loss function.
The system incorporates ASR features when generating speaker indices, increasing the likelihood of accurately assigning words to the correct speaker.
The output is a serialised transcription, including both the spoken words and the corresponding speaker indices. 
DNCASR is compared with the parallel system proposed by \citet{zhengSOTTriggered2024}, where the parallel system trains the neural clustering and ASR modules separately. 
Experimental results on a real meeting corpus show that DNCASR gives better speaker prediction than the parallel system.

The paper is organised as follows: 
\cref{sec:rwork} reviews related work, 
\cref{sec:method} presents the joint clustering and ASR system, DNCASR, together with the training and decoding pipelines,
\cref{sec:exp_setup} describes the experimental setup, 
and \cref{sec:results} provides the results, followed by the conclusions.

%% file: sections/related_work.tex
\section{Related Work}
\label{sec:rwork}

To avoid using non-neural algorithms in a speaker-attributed ASR system,
some studies extend the ASR output vocabulary to include speaker tokens.
For instance, a single neural network was proposed for transcribing doctor-patient conversations \cite{shafeyJointSpeech2019}. 
The model first outputs a speaker role and then the words spoken by that speaker. 
However, this approach is limited to only two speaker roles, doctor and patient, making it unsuitable for general multi-speaker conversations. 
Other studies can handle more than two speakers but require speaker profiles to be known in advance \cite{kandaJointSpeaker2020,luStreamingMultitalker2021}.

A parallel system proposed by \citet{zhengSOTTriggered2024} performs speaker-attributed transcription without relying on non-neural clustering.
It uses a neural clustering module to generate speaker indices and an SOT ASR to transcribe spoken words, with both components trained independently. 
The neural clustering module in the parallel system is referred to as the segment-level discriminative neural clustering (segment-level DNC), based on the original DNC proposed in \citet{liDiscriminativeNeural2020}.
The original DNC uses the Transformer encoder-decoder architecture \cite{vaswaniAttentionAll2017}, taking utterance-level speaker embeddings and requires known utterance boundaries.
The model outputs one speaker index for each utterance. 

\begin{figure}[ht]
\centering
\includegraphics[width=\linewidth]{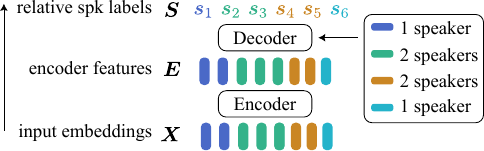}
\caption{Model architecture for segment-level DNC, colours represent different VAD segments.}
\label{fig:sdnc}
\vspace{-3mm}
\end{figure}

Segment-level DNC does not require utterance boundaries information since it uses window-level speaker embeddings.
\cref{fig:sdnc} shows the architecture of segment-level DNC, which also follows a standard Transformer encoder-decoder design. 
Here, $\bm{X}$ represents a sequence of window-level speaker embeddings. 
The encoder output, $\bm{E}$, has the same length as $\bm{X}$, but the number of output speaker indices from the decoder for each segment corresponds to the number of speakers in that segment.
In \cref{fig:sdnc}, the first segment (in blue) contains one speaker, resulting in a single index, $\bm{s}_1$, from the decoder for that segment. 
For the second segment (in green), external information indicates that there are two speakers, so the decoder produces indices $\bm{s}_2$ and $\bm{s}_3$.
The first speaker in a meeting is assigned index 0, the second assigned index 1, and so on. 
When the decoder outputs speaker indices for the current segment, its cross-attention mechanism focuses solely on the encoder features of that segment. 
To use the parallel system, a long meeting is divided into many segments, usually done by using a VAD system. 
Speaker turns within a VAD segment are merged during training to ensure they come from unique speakers.
Each VAD segment is processed by the SOT ASR to generate word tokens, including speaker change tokens. 
By counting these special tokens, segment-level DNC determines the number of speaker indices needed for each segment and generates indices for the entire meeting.
The final step involves combining the speaker indices with their corresponding word tokens from the SOT ASR.

%% file: sections/method.tex
\section{DNCASR Methodology}
\label{sec:method}

The parallel system \cite{zhengSOTTriggered2024} consists of two modules: segment-level DNC and SOT ASR (referred to as DNC and ASR).
It is challenging to jointly train these two modules since their inputs differ significantly.
Each training input to DNC module is window-level speaker embeddings of an entire meeting, while the ASR module operates on waveforms of individual VAD segments. 
Using the waveform of an entire meeting as input to the ASR module is impractical due to memory constraints.
During inference, the ASR module only provides the number of speakers in each VAD segment to the DNC module.
Since the ASR module does not share any further information about the order of the speakers, the speaker indices generated by DNC may not align with the serialised output from ASR. 
For instance, DNC might assign the speaker index of the second turn in the serialised output to the first turn.
This misalignment can result in incorrect speaker-attributed transcriptions.

\begin{figure}[ht]
\centering
\includegraphics{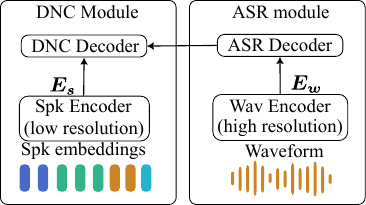}
\caption{Model architecture of DNCASR, $\bm{E_w}$ is the output of the Wav encoder, $\bm{E_s}$ is the output of the speaker encoder.}
\vspace{-2mm}
\label{fig:dncasr}
\end{figure}

This paper introduces two-stage joint fine-tuning of the DNC and ASR modules with an added link component in the DNC module, where the entire system is called DNCASR.
The link component uses cross-attention to align DNC features with the hidden features from the ASR module.
As shown in \cref{fig:dncasr}, 
the ASR module sends high-resolution information to the DNC module, enabling it to generate speaker indices that match the speaker turns identified by ASR. 
Unlike the parallel system, DNCASR does not merge speaker turns from the same speaker within a VAD segment during training, allowing non-adjacent turns to belong to the same speaker.

\subsection{Joint Fine-tuning -- Stage 1}

In DNCASR, the DNC and ASR modules are initially pre-trained separately before entering the first joint training stage. During this joint-training stage, the DNC module is trained to generate the speaker indices from the first segment to the current segment, while the ASR module is trained to produce the word sequence for the current segment.
The loss function for joint training is the sum of the loss functions for the DNC and ASR modules, consisting of two cross-entropy losses.

\begin{figure}[ht]
\centering
\includegraphics[width=\linewidth]{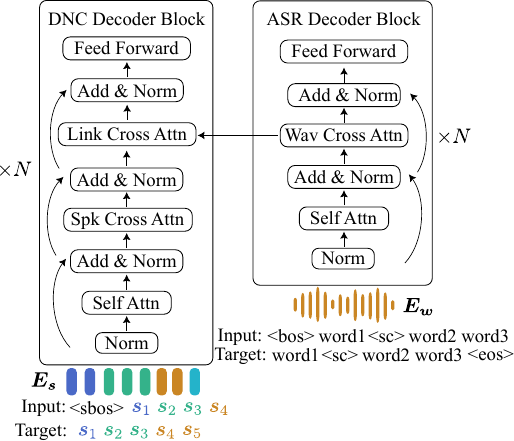}
\caption{Detailed architecture of the decoder blocks in DNCASR during joint fine-tuning stage 1. $\text{<bos>}$ and $\text{<sbos>}$ indicate start of sequence tokens.}
\label{fig:dncasr_stage1_layer}
\vspace{-2mm}
\end{figure}

\cref{fig:dncasr_stage1_layer} shows the architecture of the decoder blocks in DNCASR for this stage, where $N$ is the number of blocks in the decoder. 
The output of the final DNC decoder block is projected to generate target speaker indices from the first segment to the current segment. 
The output of the final ASR decoder block is projected to generate word tokens of the current segment, including speaker change tokens ($\text{<sc>}$) and an end-of-sequence token ($\text{<eos>}$). 

The ASR decoder block is a standard Transformer decoder \cite{vaswaniAttentionAll2017}, which has one self-attention module (Self Attn) and one cross-attention module (Wav Cross Attn).
The output of the Wav Cross Attn module in each block is referred to as $\bm{W}_\mathrm{CA}$.
In joint fine-tuning stage 1, $\bm{W}_\mathrm{CA}$ of the $n$-th block is used in the Link Cross Attn module in the $n$-th block of the DNC decoder.

To link the DNC and ASR decoder blocks, a modified Transformer decoder block is proposed for the DNC decoder. 
Each DNC decoder block now has two cross-attention modules: the first cross-attention module (Spk Cross Attn) attends to the speaker encoder outputs $\bm{E_s}$:
\begin{equation}
\begin{aligned}[b]
\bm{S}_\mathrm{CA}[i] &= \mathrm{CA}(\bm{Q}, \bm{K}, \bm{V}) = \mathrm{CA}(\bm{Q}, \bm{K}) \\
&= \mathrm{CA}(\bm{S}_\mathrm{SA}[i], \bm{E_s}\odot \mathrm{mask}_s[i])
\end{aligned}
\label{eq:sca}
\end{equation}
where $\bm{S}_\mathrm{CA}[i]$ is the output of the Spk Cross Attn module for the $i$-th target speaker index of the entire meeting, $\mathrm{CA}$ represent cross attention function with query ($\bm{Q}$), key ($\bm{K}$) and value ($\bm{V}$) matrices. Since $\bm{K}$ and $\bm{V}$ are identical in \cref{eq:sca}, we omit $\bm{V}$ in the equations. 
$\bm{S}_\mathrm{SA}[i]$ is the output of the self-attention module for the $i$-th target speaker index.
$\odot$ represents element-wise multiplication, and $\mathrm{mask}_s[i]$ is a mask matrix that is used to ignore the Spk encoder outputs $\bm{E}_s$ of all segments except the one corresponding to the $i$-th target speaker index.

The second cross-attention module (Link Cross Attn) in the $n$-th DNC decoder block attends to the output $\bm{W}_\mathrm{CA}$ from the Wav Cross Attn module in the $n$-th ASR decoder block:
\begin{equation}
\begin{aligned}
&\bm{L}_\mathrm{CA}[i] = \mathrm{CA}(\bm{S}_\mathrm{CA}[i], \bm{W}_\mathrm{CA}\odot \mathrm{mask}_l[i])
\end{aligned}
\label{eq:lca}
\end{equation}
where $\bm{L}_\mathrm{CA}[i]$ is the output of the Link Cross Attn module for the $i$-th target speaker index of the entire meeting, $\bm{S}_\mathrm{CA}[i]$ is the output of the Spk Cross Attn module for the same target speaker index. 
Each target output word token, including <sc> and <eos>, has its own $\bm{W}_\mathrm{CA}$ in each ASR decoder block.
$\bm{W}_\mathrm{CA} \odot \mathrm{mask}_l[i]$ serves as the $\bm{K}$ and $\bm{V}$ for the Link Cross Attn module, where $\mathrm{mask}_l[i]$ masks out the $\bm{W}_\mathrm{CA}$ features of all word tokens except those corresponding to the $i$-th target speaker index.
\begin{figure}[h!]
\centering
\includegraphics{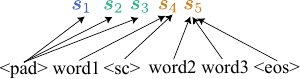}
\vspace{-2mm}
\caption{Information flow from $\bm{W}_\mathrm{CA}$ to $\bm{S}_\mathrm{CA}$ features between each pairs of decoder block in stage 1.}
\vspace{-2mm}
\label{fig:dncasr_stage1_link}
\end{figure}
To further explain \cref{eq:lca}, the information flow between the corresponding ASR and DNC decoder blocks is shown in \cref{fig:dncasr_stage1_link}.
$\bm{s}_1$ to $\bm{s}_3$ are the speaker indices of the past segments, while $\bm{s}_4$ and $\bm{s}_5$ are the speaker indices in the current segment.
Since the ASR module processes only the current segment, only the speaker indices in that segment are aligned with the word token features. 
Therefore, for $\bm{S}_\mathrm{CA}$ features corresponding to past segments, they attend to a learnable embedding, denoted as $\text{<pad>}$, while the $\bm{S}_\mathrm{CA}$ features in the current segment attend to the $\bm{W}_\mathrm{CA}$ features of the word tokens in the corresponding speaker turn. 
For example, $\bm{s}_4$ attends to $\bm{W}_\mathrm{CA}$ features corresponding to the ASR output tokens $\text{word1}$ and $\text{<sc>}$, $\bm{s}_5$ attends to $\bm{W}_\mathrm{CA}$ features corresponding to $\text{word2}$, $\text{word3}$ and $\text{<eos>}$. 
Here we align the $\bm{W}_\mathrm{CA}$ features corresponding to the $\text{<sc>}$ and $\text{<eos>}$ tokens to the speaker turn on the left. 
In this stage, the training targets are the speaker indices from the first segment up to the current segment, together with the word tokens of the current segment. 

\subsection{Joint Fine-tuning -- Stage 2}

In the second stage of joint training, only the DNC decoder is fine-tuned, using pre-computed $\bm{W}_\mathrm{CA}$ features for all speaker turns throughout the entire meeting. 
Each DNC decoder block receives the corresponding $\bm{W}_\mathrm{CA}$ features for all word tokens across all segments.
The architecture of DNCASR remain the same as in \cref{fig:dncasr_stage1_layer}, but the ASR module is frozen in this stage. 

\begin{figure}[ht]
\centering
\includegraphics{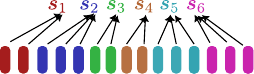}
\caption{Information flow in stage 2. Colours indicate different speaker turns, with each coloured rectangle representing a $\bm{W}_\mathrm{CA}$ feature.}
\label{fig:dncasr_stage2_link}
\vspace{-3mm}
\end{figure}

As shown in \cref{fig:dncasr_stage2_link}, after computing all $\bm{W}_\mathrm{CA}$ features for the entire meeting in advance, each speaker index now attends to its own $\bm{W}_\mathrm{CA}$ features in each DNC decoder block, rather than attending to a learnable $\text{<pad>}$ embedding as in stage 1.
In stage 2, the training target is the speaker indices of the entire meeting. 

\subsection{Inference}
The inference procedure for DNCASR is similar to the parallel system, where the ASR module first decodes all VAD segments, and then the DNC module decodes speaker labels for the entire meeting. 
The difference is that in DNCASR, the $\bm{W}_\mathrm{CA}$ features of all VAD segments are stored, allowing the DNC module to attend to them during decoding.
The inference procedures differ between stage 1 and stage 2: stage 1 requires running DNC separately for each segment, whereas stage 2 allows DNC to be applied to all segments in a single pass.
More detail is given in \autoref{app:decoding_details}.

\subsection{Constrained Diaconis Augmentation}
\label{sec:diaconis}

Training the DNC module requires a large amount of meeting data. However, publicly available real meeting datasets, such as AMI, 
are limited in size. 
To ensure sufficient training data and avoid overfitting, Diaconis Augmentation is used \cite{liDiscriminativeNeural2020} to apply random rotations \cite{stewartEfficientGenerationRandom1980,diaconis1987} to the speaker embeddings, thereby increasing the amount of training data.
However, excessive rotation of the speaker embeddings may lead to a decrease in performance. 
In this paper, we propose a Constrained Diaconis Augmentation (CDA) method to control the rotation angle of the speaker embeddings. 
More detail of the formulation is given in \cref{sec:diaconis_formula}. 

%% file: sections/exp_setup.tex
\section{Experimental Setup}
\label{sec:exp_setup}

This section describes two datasets used for training and evaluating DNCASR,
along with the model configuration, pre-training data and evaluation metrics. 

\subsection{Synthetic Data}
\label{sec:synthetic_data}
To test the proposed DNCASR in a controlled setting, 
synthetic meeting data was generated from LibriSpeech \cite{panayotovLibrispeechASR2015}. 
For training, 6,000 simulated meetings were created, each with 8 speakers, 
and each simulated meeting was approximately 10 minutes long. 
Each segment consists of a random selection of 1 to 5 utterances from the LibriSpeech \texttt{train\_960h} set, where adjacent utterances always come from different speakers.
Adjacent utterances can have up to a 25\% overlap ratio, while the total overlap ratio of the entire meeting is around 5\%.
For evaluation, 20 simulated meetings were created in the same manner as the training set, but using utterances from the LibriSpeech \texttt{dev\_clean} set.
The training data is used to train the SOT ASR system, fine-tune the SDNC in the parallel system, and perform joint fine-tuning for DNCASR.

\subsection{AMI}
\label{sec:ami_data}
AMI \cite{carlettaAmiMeeting2006} is a real meeting dataset that contains 100 hours of meeting recordings with 3 to 5 speakers in each meeting. 
In this paper, only the multi-distant microphone (MDM) audio, beam-formed using BeamformIt \cite{angueraAcousticBeamforming2007}, is used for the AMI experiments.
There are 135 meetings for the training set, 18 for the Dev set and 16 for the Eval set. 
The manual segmentation has been found to label a lot of non-speech silence regions as speech \cite{sunCombinationDeep2021}, the same procedure in \citet{sunCombinationDeep2021} was used to remove non-speech regions with a pre-existing HMM-based ASR system \cite{youngHTKBook2015} to do force alignment. 
Compared to the original speech regions, the silence-stripped data reduces the total duration by 9.9\% for Dev and by 11.7\% for Eval.

Since the AMI dataset is limited in size, data augmentation methods in \citet{zhengSOTTriggered2024} were applied to expand the training data. 
This includes Diaconis augmentation, VAD segment permutation, and gradually increasing the length of the meetings used during training.

\subsection{Model Configuration}

The Spk encoder, DNC decoder, and ASR decoder in DNCASR were trained from scratch, which all used 6-layer Transformer-based decoder with 4 attention heads. 
The hidden size is set to 256, and the feed-forward size is set to 2048.
The Wav encoder uses WavLM \cite{chenWavLMLargescale2022}, a foundation model pre-trained using a self-supervised learning (SSL) approach.
The WavLM used in this paper is the wavlm-base-plus model from \citet{wolfTransformersStateoftheart2020}. 
The window level speaker embeddings are extracted with a frozen ECAPA-TDNN \cite{dawalatabadECAPATDNNEmbeddings2021} model, which was trained on VoxCeleb \cite{nagraniVoxcelebLargescale2020} and VoxCeleb2 \cite{chungVoxCeleb2Deep2018} datasets. 
The window size is 1.5s with stride of 0.5s. 
More details can be found in \cref{app:model_details}.

For experiments on the simulated data, the supervised data is derived from synthetic mixtures generated using the LibriSpeech dataset.
For AMI experiments, the supervised data comes exclusively from the AMI-MDM data.

\subsection{Pre-training DNC and ASR Modules}

The data described in \cref{sec:synthetic_data,sec:ami_data} are referred to as VAD segment data. 
When pre-training the DNC and ASR modules separately,
the ASR module uses VAD segment data, while
the DNC module needs to be pre-trained on segments that contain only one speaker \cite{zhengSOTTriggered2024}. 
More detail for DNC pre-training is given in \cref{app:pre-train-DNC}. 

\subsubsection{Synthetic Data}
\label{sec:synthetic_pretrain}
To pre-train the DNC module, a separate dataset is used to generate 22.5k simulated meetings from Librispeech \texttt{train\_960h}, where each segment contains only one speaker. The other settings are the same as those in the synthetic data training set.

\subsubsection{AMI}
\label{sec:ami_pretrain}
\begin{figure}[ht]
\centering
\includegraphics{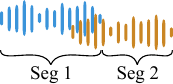}
\caption{A single overlapping VAD segment split into two segments using the FSS method.}
\label{fig:fss}
\vspace{-3mm}
\end{figure}

To pre-train the DNC module without generating synthetic meeting data, we ensure one segment per meeting by applying the First Speaker Segmentation (FSS) method from \citet{zhengSOTTriggered2024} to the AMI-MDM VAD segment data. This method splits overlapping speech segments into multiple segments, each assigned to a single speaker.
FSS assigns the overlap region to the speaker who speaks first, then splits the segment at the end of the overlap region, as shown in Figure \ref{fig:fss}.

\subsection{Evaluation Metrics}

Three metrics are used for evaluation: diarisation error rate (DER), word error rate (WER), and concatenated minimum-permutation word error rate (cpWER). 
DER is a time-based metric, while cpWER and WER are word-based metrics. 
When scoring DER, a 0.25s collar is used, and scoring includes overlap regions. 
Since we use VAD segments with oracle boundaries, the DER is the same as the speaker error rate. 
MeetEval \cite{vonneumannMeetEvalToolkit2023} is used to calculate cpWER, where cpWER concatenates transcriptions from the same speaker and finds the minimum WER across all possible speaker mappings between the predicted speaker indices and reference speaker IDs. 

%% file: sections/results.tex
\section{Experimental Results}
\label{sec:results}

The section first shows the results on the synthetic data, then the results on the AMI-MDM dataset. 

\subsection{Synthetic Data Experiments}

\begin{table}[ht]
\begin{center}
\begin{tabular}{cccc}
\toprule[1.5pt]
Model & WER & cpWER\\
\midrule
Parallel & 3.5 & 13.4 \\
\midrule
DNCASR (S1) & 3.5 & 9.5 \\
DNCASR (S2) & 3.5 & 8.7 \\
\bottomrule
\end{tabular}
\caption{\%WER and \%cpWER on the synthetic data. S1 and S2 refer to the first and second joint fine-tuning stages of the DNCASR system.}
\label{tab:syn_freeze_asr}
\end{center}
\vspace{-3mm}
\end{table}

The DNC and ASR modules were separately pre-trained: DNC was pre-trained on synthetic meetings without overlapping speech (\cref{sec:synthetic_pretrain}), while ASR was pre-trained on VAD segment data (\cref{sec:synthetic_data}).
The pre-trained modules are used to initialise the parallel and DNCASR systems. 
The parallel system was re-implemented following  \citet{zhengSOTTriggered2024}, where the DNC module is fine-tuned on VAD segment data after pre-training.
The DNC module in DNCASR was fine-tuned on VAD segment data together with the ASR module. 
\cref{tab:syn_freeze_asr} shows the results of the parallel and DNCASR systems on the synthetic data when the ASR module is frozen after pre-training.
Unlike the DNC module in the parallel system, which does not use hidden features from the ASR module, the DNC module in DNCASR (stages 1 and 2) uses these hidden features.
The results demonstrate that using the hidden features of word tokens can improve DNC performance when the ASR modules are the same across the three different setups.
Both stages 1 and 2 fine-tuning of DNCASR outperform the parallel system, reducing cpWER by 29.1\% (stage 1) and 35.1\% (stage 2). 
Extra results using the same checkpoint to evaluate on test sets with variable number of speakers can be found in \cref{app:var_speaker_num}, and the LibriCSS \cite{chenContinuousSpeech2020} OV10 results can be found in \cref{app:libricss_ov10}.

\begin{table}[ht]
\begin{center}
\begin{tabular}{ccccc}
\toprule[1.5pt]
Model & DER & cpWER \\
\midrule
Parallel & 3.9 & 6.8 \\
\midrule
DNCASR (S1) & 2.7 & 4.2 \\
DNCASR (S2) & 1.6 & 3.6 \\
\bottomrule
\end{tabular}
\caption{\%DER and \%cpWER on the synthetic data with oracle word sequence and speaker turns.}
\label{tab:syn_oracle}
\end{center}
\vspace{-3mm}
\end{table}

\cref{tab:syn_oracle} shows the results of the DNCASR system on the synthetic data when using oracle word sequence. 
In both the parallel and DNCASR systems, the DNC module is provided with the oracle speaker turns for each segment. 
In the DNCASR system, the ASR module uses the oracle word sequence to provide the hidden features of word tokens to the DNC module.
The DERs in \cref{tab:syn_oracle} show the performance of the DNC module when the number of speaker turns is known. 
The results indicate that the DNCASR system outperforms the parallel system, reducing the DER by 30.8\% in stage 1 and 59.0\% in stage 2.
The cpWER in \cref{tab:syn_oracle} shows the cpWER when the errors are solely due to speaker assignment.
The cpWER is reduced by 38.2\% in stage 1 and 47.1\% in stage 2 compared to the parallel system.

\subsection{AMI Experiments}

\begin{table}[ht]
\begin{center}
\begin{tabular}{ccccc}
\toprule[1.5pt]
Model & DER (FSS) & WER \\
\midrule
Cascaded & 5.4/4.0 & 22.4/24.7 \\
\midrule
\makecell{Parallel (Pretrain) \\ \citet{zhengSOTTriggered2024}} & 
5.6/4.4 & 25.8/26.6 \\
\midrule
DNCASR (Pretrain) & 5.3/4.1 & 25.1/27.1 \\
\midrule
DNCASR (S1) & - & 24.9/26.6 \\
\bottomrule
\end{tabular}
\caption{\%DER (score overlap regions with 0.25 collar) and \%WER on Dev/Eval set of AMI-MDM. \%WER for the cascaded system is computed per utterance; for other systems, it is per VAD segment.}
\label{tab:ami_fss}
\end{center}
\vspace{-4mm}
\end{table}

For the AMI experiments, the ASR module was pre-trained on the AMI MDM VAD segment data, while the DNC module was pre-trained on the FSS segment data, as described in \cref{sec:ami_pretrain}.
With FSS segments, although neither the parallel nor the DNCASR system outputs time information, both can use the boundaries of the FSS segments to calculate the DER.
The DNCASR system is compared to the parallel system pre-training results from \citet{zhengSOTTriggered2024}, which also uses the FSS segments for pre-training the DNC.
Another baseline is the cascaded system, which uses the same window-level embeddings as DNCASR. 
This approach begins by applying spectral clustering to produce speaker-homogeneous segments, which are then decoded using an ASR model trained on individual utterances. 
Unlike the SOT ASR, which is trained using oracle VAD segmentations, the ASR model in the cascaded system is trained on single utterances.
The DER results show that the DNCASR pre-training results are comparable to those of the cascaded and parallel systems.
Although the WER performance on the Eval set is slightly worse than the parallel system during pre-training, it matches the parallel system's performance after joint fine-tuning.
After fine-tuning without FSS segments, the DNCASR system lacks accurate start and end times for each predicted speaker turn within a VAD segment, making it unsuitable for calculating DER.

\begin{table}[ht]
\begin{center}
\begin{tabular}{ccccc}
\toprule[1.5pt]
\multirow{2}{*}{Model} & \multirow{2}{*}{cpWER} & cpWER \\
& & (Multi) \\
\midrule
Cascaded & 35.2/33.0 & 46.0/46.1 \\
\midrule
\makecell{Parallel \\ \citet{zhengSOTTriggered2024}} & 34.8/34.6 & 49.8/49.2 \\
\midrule
DNCASR (S1) & 33.2/34.7 & 47.3/49.5 \\
\midrule
DNCASR (S2) & 31.3/32.1 & 43.4/44.8 \\
+ CDA & 30.7/31.5 & 42.5/44.1 \\
\bottomrule
\end{tabular}
\caption{\%cpWER on AMI Dev/Eval set. \%cpWER-Multi is the \%cpWER of multi-talker segments. }
\label{tab:ami_cpwer}
\end{center}
\vspace{-4mm}
\end{table}

\cref{tab:ami_cpwer} shows the cpWER results on the AMI-MDM dataset after fine-tuning the DNC module on the training set.
Stage 1 fine-tuning of the DNCASR system (S1) yields better cpWER performance than both the cascaded and parallel systems on the Dev set, with relative reductions of 5.7\% and 4.6\% respectively. 
However, it performs slightly worse than the two baselines on the Eval set.
For cpWER-Multi, which is the cpWER on the multi-talker VAD segments in the reference, both the parallel and DNCASR (S1) are worse than the cascaded system. 
Since the AMI dataset has more segments and a longer total duration than the synthetic data, omitting or restricting hidden features for word tokens to the current segment may be insufficient for improving speaker assignment.
After stage 2 fine-tuning, the DNCASR system outperforms both the cascaded and parallel systems on the Dev and Eval sets, achieving relative cpWER reductions of 11.1\% and 2.7\% compared to the cascaded system, and 10.1\% and 7.2\% compared to the parallel system.
When CDA is applied, the scale is set randomly between 0 and 10 for each training example.
After further fine-tuning the DNCASR (S2) system with CDA, cpWER is further reduced, with relative improvements of 12.8\% and 4.5\% over the cascaded system, and 11.8\% and 9.0\% over the parallel system on the Dev and Eval sets, respectively.
The cpWER-Multi is reduced by 14.7\% and 10.4\% on the Dev and Eval sets compared to the parallel system, indicating that the majority of the improvement comes from the multi-talker segments. 
Some DNCASR outputs are shown in \cref{sec:example_outputs}.
\cref{app:wavlm-large} presents results comparing the best DNCASR setups using wavlm-base-plus and wavlm-large, which leads to more than 10\% relative cpWER reductions in both AMI Dev/Eval by using the larger SSL model. 

\begin{table}[ht]
\begin{center}
\begin{tabular}{ccccc}
\toprule[1.5pt]
\multirow{2}{*}{Model} & \multirow{2}{*}{DER} & \multicolumn{3}{c}{cpWER} \\
& & All & Single & Multi \\
\midrule
DNCASR (S1) & 6.7 & 19.3 & 5.6 & 33.3 \\
\midrule
DNCASR (S2) & 6.5 & 17.8 & 6.5 & 28.5 \\
+ CDA & 6.3 & 17.4 & 6.3 & 28.3 \\
\bottomrule
\end{tabular}
\caption{\%DER and \%cpWER on the AMI Eval set with oracle words. `all' refers to all segments, Single and multi refer to single- and multi-talker segments.}
\label{tab:ami_oracle}
\end{center}
\vspace{-4mm}
\end{table}

\cref{tab:ami_oracle} shows the results of the DNCASR system on the AMI-MDM Eval set when using the oracle word sequence. 
Given the oracle speaker turns in each segment, the DER was computed using the oracle timestamps for each turn.
There is a consistent improvement in DER from S1 to S2+CDA, with a relative reduction of 6.0\%, and cpWER shows an overall 9.8\% relative reduction. 
However, the cpWER on the single-talker segments is worse in stage 2 than in stage 1, indicating that the improvement in cpWER primarily comes from the multi-talker segments, where a relative reduction of 15.0\% is observed.

\begin{table}[ht]
\begin{center}
\begin{tabular}{ccccc}
\toprule[1.5pt]
Model & p-value & \#meetings improved \\
\midrule
DNCASR (S2) & 1.1E-6 & 30 \\
+ CDA & 5.1E-9 & 31\\
\bottomrule
\end{tabular}
\caption{cpWER Wilcoxon signed-rank test results on the AMI Dev and Eval compared against DNCASR (S1). 
In total there are 34 meetings. 
}
\label{tab:ami_sign}
\end{center}
\vspace{-4mm}
\end{table}

\cref{tab:ami_sign} shows the single sided Wilcoxon signed-rank test \cite{wilcoxonIndividualComparisons1945} results for the cpWER on the AMI-MDM set between stage 1 and stage 2 of the DNCASR system. 
The cpWER of individual meeting pairs is compared, and the p-value is calculated to determine whether the cpWER of stage 2 is significantly lower than that of stage 1. 
The cpWERs of all 34 meetings in the combined AMI Dev and Eval set are shown in \cref{sec:indiv_cpwer}. 
\cref{tab:ami_sign} shows that DNCASR (S2) has 30 meetings with a lower cpWER than DNCASR (S1), and DNCASR (S2+CDA) has 31 meetings with a lower cpWER than DNCASR (S1). 
The p-values show that the improvements in cpWER from stage 1 to stage 2 are highly statistically significant.

%% file: sections/conclusion.tex
\section{Conclusions}
\label{sec:conclusions}

This paper proposes a joint neural clustering and ASR system that allows end-to-end joint training.
By incorporating the ASR hidden features into the neural clustering module, the system is able to predict more accurate speaker indices in overlapping segments.
We also introduce the Constrained Diaconis Augmentation method to maintain the rotated speaker embeddings close to their original values, further enhancing the accuracy of predicted speaker indices.
The best DNCASR system outperforms the parallel system on both the AMI Dev and Eval sets, achieving a relative cpWER reduction of 11.8\% and 9.0\% respectively, as well as a 14.7\% and 10.4\% relative reduction on multi-talker segments, highlighting the effectiveness of DNCASR in handling complex conversational scenarios.

%% file: sections/appendix.tex
\section{Model Details}
\label{app:model_details}
The DNCASR model consists of four main components: the Wav encoder, the Spk encoder, the DNC decoder, and the ASR decoder.
In total the number of parameters in the entire DNCASR system is 117M, where most of the parameters are in the Wav encoder, which is 94M. 
The Wav encoder is a pre-trained WavLM model (wavlm-base-plus) from \citet{wolfTransformersStateoftheart2020}. 
All experiments were conducted on a single A100 GPU with 80GB of memory. 
For the real-world AMI data experiments, the ASR module pre-training took approximately 25 minutes per epoch, with a total of 60 epochs completed in around 25 GPU hours. 
The DNC module was pre-trained on FSS segments for 250 epochs, with each epoch taking 5 minutes, amounting to a total of 21 GPU hours. 
Fine-tuning the DNCASR model was carried out in two stages. 
Stage 1 took 2 hours per epoch for 10 epochs, requiring about 20 GPU hours in total. 
Stage 2 took 1.5 hours per epoch for 5 epochs, followed by 3 additional epochs with Constrained Diaconis Augmentation, resulting in a total of 12 GPU hours.
The learning rate was set to 5E-4 with Adam optimizer with linear warm-up for the first 20\% of the training steps. 

\section{Inference Details}
\label{app:decoding_details}

\begin{figure}[ht]
\centering
\includegraphics{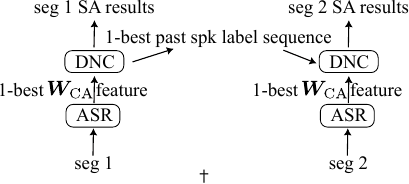}
\caption{Illustration of DNCASR's stage 1 decoding.}
\label{fig:cha7:stage1_decoding}
\end{figure}

The inference procedure for DNCASR fine-tuning stage 1, illustrated in \cref{fig:cha7:stage1_decoding}, involves executing both the DNC and ASR modules once for each segment. 
Inference begins with the ASR module, which uses beam search to generate serialised output word tokens, including the special tokens <sc> and <eos>, for each segment. 
The top-1 (1-best) word tokens from the current segment are then used to produce the corresponding 1-best $\bm{W}_\mathrm{CA}$ features for each ASR decoder block.

Subsequently, the DNC module decodes the speaker labels for the current segment. 
The 1-best speaker label sequence from previous segments is provided as context to the DNC decoder to decode the speaker labels of the current segment. 
Therefore, beam search is applied only to the current segment, and the 1-best speaker labels from this segment are appended to update the 1 best sequence of past speaker labels.
During inference, outputs associated with past speaker labels (context) attend to the <pad> embeddings, whereas outputs corresponding to the current segment's speaker labels attend to this segment's 1-best $\bm{W}_\mathrm{CA}$ features.

\begin{figure}[ht]
\centering
\includegraphics{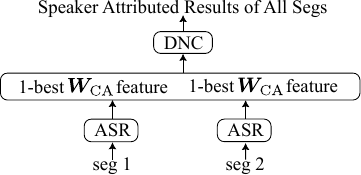}
\vspace{-1mm}
\caption{Decoding illustration of DNCASR.}
\label{fig:decode}
\vspace{-2mm}
\end{figure}

The inference procedure for DNCASR fine-tuning stage 2 is shown in \cref{fig:decode}. 
The ASR module first decodes each VAD segment using beam search to generate word tokens, saving the 1-best word tokens along with the corresponding $\bm{W}_\mathrm{CA}$ features. 
The DNC module then decodes the speaker labels for the entire meeting, attending to the $\bm{W}_\mathrm{CA}$ features of the corresponding speaker turn.

\section{Pre-training DNC}
\label{app:pre-train-DNC}
During pre-training of the DNC, the Link Cross Attention module is absent and will be added and trained from scratch during the fine-tuning stages. 
Length scheduling plays a crucial role in pre-training the DNC module, starting with shorter meeting segments and gradually increasing the segment length until the full meeting is covered.
Since DNC and ASR pre-training can be performed simultaneously and takes roughly the same amount of time, it is more efficient to pre-train both modules before proceeding to jointly fine-tune. We have tested that during fine-tuning stage 1, with the same stage 1 fine-tuning procedure, if the DNC is not pre-trained, it is unable to learn how to provide speaker indices at all.

% \newpage
\section{Constrained Diaconis Augmentation (CDA) Formula}
\label{sec:diaconis_formula}

The transformation formula in \citet{stewartEfficientGenerationRandom1980} states that the transformation $\bm{H} \in \mathbb{R}^n$ can be constructed as follows:
\begin{equation}
\begin{aligned}[b]
\bm{H} = \bm{D}\bm{H}_1\bm{H}_2\cdots \bm{H}_{n-1}
\end{aligned}
\label{eqn:diaconis}
\end{equation}
where the term $H_j$ is defined as 
\[
    \left(
    \begin{array}{ccccc}
    \bm{I} & 0\\
    0 & \bm{\bar{H}_j} 
    \end{array}
    \right)
\]  
and $\bm{D}$ is a diagonal sign matrix s.t. $diag(\bm{D})[j-1] = sign(diag(\bm{\bar{H}}_j)[0])$. Each $\bar{H}_j$ is a Householder transformation matrix in $\mathbb{R}^{n-j+1}$:
\begin{equation}
\begin{aligned}[b]
\bm{\bar{H}}_j = \bm{I}_j - 2\frac{\bm{v}_j\bm{v}_j^T}{\bm{v}_j^T\bm{v}_j}
\end{aligned}
\label{eqn:householder}
\end{equation}
where $\bm{v}_j \in \mathbb{R}^{n-j+1}$ is a random vector. 
To control the transformation angle, we only need to control the first element of the random vector $\bm{v}$:
\begin{equation}
\begin{aligned}[b]
\bm{\tilde{v}}_j = [\bm{v}_j[0]-\mathrm{scale}, \bm{v}_j[1], \cdots]
\end{aligned}
\label{eqn:control}
\end{equation}
where $\mathrm{scale}$ is a constant value that controls the rotation angle.
When the scale is set to $\infty$, $\bm{\bar{H}}_j$ becomes an identity matrix, except for the first element, which is -1. 
In this case, $\bm{D}\prod_{j=1}^{n-1}\bm{H}_j = \bm{I}$.
Setting scale to be 0 is equivalent to the original unconstrained Diaconis Augmentation. 

Applying rotations to 100 32-d random vectors, each with a different $\bm{H}$, produces the averaged absolute rotation angles shown in \cref{tab:diaconis_angle}.
\begin{table}[ht]
\begin{center}
\begin{tabular}{c|cccccccc}
\toprule[1.5pt]
& C=0 & C=1 & C=10 & C=100 \\
\midrule
Angle & 90.0 & 83.2 & 29.6 & 3.2 \\ 
\bottomrule
\end{tabular}
\caption{Averaged absolute rotation angles with different constrained scales $C$.}
\label{tab:diaconis_angle}
\end{center}
\vspace{-3mm}
\end{table}

\section{Additional Results}

\subsection{Variable Number of Speakers}
\label{app:var_speaker_num}

To evaluate the ability of DNCASR to handle varying numbers of speakers, we use the same DNCASR (S2) checkpoint from \cref{tab:syn_freeze_asr} to test on datasets with different speaker counts. 
Specifically, we follow the same procedure to generate two additional test sets containing 7 and 6 speakers, respectively.
\begin{table}[ht]
\begin{center}
\begin{tabular}{cc}
\toprule[1.5pt]
\# speaker & cpWER \\
\midrule
8 & 8.7 \\
7 & 7.2 \\
6 & 10.2 \\
\bottomrule
\end{tabular}
\caption{Using DNCASR (S2) in \cref{tab:syn_freeze_asr} on synthetic meetings with different number of speakers per meeting. }
\label{tab:syn_diff_speaker}
\end{center}
\end{table}
Surprisingly, despite being trained solely on 8-speaker data, our model performed well in 7-speaker scenarios. In 20 simulated 7-speaker meetings, the model correctly identified the correct number of speakers in 10 of the meetings. For the remaining 10 meetings, while the model slightly overestimated the number of speakers (assigning 8), the 8th speaker index only received a small number of speaker turns, which had minimal impact on the overall results. In the 6-speaker scenario, the cpWER increases noticeably, indicating that the model indeed needs to be trained with varying numbers of speakers to effectively handle situations with a more diverse number of speakers.

\subsection{LibriCSS OV10}
\label{app:libricss_ov10}

LibriCSS \cite{chenContinuousSpeech2020} is a synthetic data created by mixing utterances from LibriSpeech \cite{panayotovLibrispeechASR2015} \texttt{test\_clean} set to create conversations.
Each conversation has 8 speakers lasting for around 10 minutes. 
The conversations are divided into sessions based on different overlap ratios. 
The anechoic (rather than replayed) OV10 session of LibriCSS most closely matches the setup of our synthetic data experiments, although conversations in OV10 have an approximate 10\% overlap ratio compared to 5\% in our experiments. 
The same DNCASR checkpoints from \cref{tab:syn_freeze_asr} are then used to evaluate the 10 conversations in this anechoic LibriCSS session. Similar to our synthetic experiments, the results on OV10 also show consistent improvement from stage 1 to stage 2 of the fine-tuning process.

\begin{table}[ht]
\begin{center}
\begin{tabular}{ccc}
\toprule[1.5pt]
\multirow{2}{*}{Model} & \%cpWER & \%cpWER \\
& (Ours) & (OV10) \\
\midrule
DNCASR (S1) & 9.5 & 11.9 \\
DNCASR (S2) & 8.7 & 7.6 \\
\bottomrule
\end{tabular}
\caption{Comparing the \%cpWER in \cref{tab:syn_freeze_asr} (ours) and in anechoic LibriCSS OV10 session using the same DNCASR checkpoints. }
\label{tab:syn_diff_speaker}
\end{center}
\end{table}

\section{Example Outputs of DNCASR}
\label{sec:example_outputs}

The following examples show the inference results of some VAD segments with multiple speaker turns within long meetings. 
\\
\\
\textbf{Example 1:}
This example shows that DNCASR S1 predicts the same speaker indices for two turns, while DNCASR S2+CDA predicts correct speaker indices.
\vspace{3mm}
\\
Reference: 
\\
\textbf{<spk1>} THE SLIDE PAD \textbf{<spk0>} WE ALSO DON'T KNOW HOW MANY BUTTONS ARE REQUIRED OR
\\
\\
DNCASR S1: 
\\
\textbf{<spk0>} THE SPICE YEAH \textbf{<spk0>} WE ALSO DON'T KNOW HOW MANY BUTTONS ARE REQUIRED OR
\vspace{3mm}
\\
DNCASR S2+CDA: 
\\
\textbf{<spk1>} THE SPICE YEAH \textbf{<spk0>} WE ALSO DON'T KNOW HOW MANY BUTTONS ARE REQUIRED OR
\\
\\
\\
\textbf{Example 2:}
This example shows that DNCASR S1 predicts the second speaker index incorrectly, while DNCASR S2+CDA predicts the correct speaker indices.
\vspace{3mm}
\\
Reference: 
\\
\textbf{<spk0>} WE HAVE ONE FOR THE ZERO AND ONE FOR THE \textbf{<spk3>} UH YEAH
\vspace{3mm}
\\
DNCASR S1: 
\\
\textbf{<spk0>} WE HAVE ONE FOR THE ZERO AND ONE FOR THE \textbf{<spk1>} YEAH
\vspace{3mm}
\\
DNCASR S2+CDA: 
\\
\textbf{<spk0>} WE HAVE ONE FOR THE ZERO AND ONE FOR THE \textbf{<spk3>} YEAH
\\
\newpage
\noindent\textbf{Example 3:}
This example shows that DNCASR S1 predicts the last two speaker turns incorrectly, while DNCASR S2+CDA predicts the correct speaker indices.
\vspace{3mm}
\\
Reference: 
\\
\textbf{<spk0>} INTRO YEAH \textbf{<spk2>} TO DO IT BECAUSE IT'S ONLY TWELVE AND A HALF EUROS YOU HAVE TO SPEND ON EVERY REMOTE CONTROL \textbf{<spk0>} YEAH THAT'S THE PROBLEM THAT'S THE MAIN PROBLEM \textbf{<spk1>} WELL I GOT F ALSO AN EMAIL FROM
\vspace{3mm}
\\
DNCASR S1: 
\\
\textbf{<spk0>} THAT'S TRUE \textbf{<spk2>} TO DO IT BECAUSE IT'S ONLY TWELVE AND A HALF EUROS YOU HAVE TO SPEND A REMOTE CONTROL \textbf{<spk1>} YEAH WELL THE PROGRAMME JUST \textbf{<spk3>} I GOT ALSO AN EMAIL FROM
\vspace{3mm}
\\
DNCASR S2+CDA: 
\\
\textbf{<spk0>} THAT'S TRUE \textbf{<spk2>} TO DO IT BECAUSE IT'S ONLY TWELVE AND A HALF EUROS YOU HAVE TO SPEND A REMOTE CONTROL \textbf{<spk0>} YEAH WELL THE PROGRAMME JUST \textbf{<spk1>} I GOT ALSO AN EMAIL FROM
\\
\\
\textbf{Example 4:}
This example shows that ASR predicts one less speaker turn than the reference, DNCASR S1 predicts the first speaker index correctly but the second speaker index incorrectly, while DNCASR S2+CDA predicts the correct speaker indices.
\vspace{3mm}
\\
Reference: 
\\
\textbf{<spk1>} YEAH BUT IT YOU CAN'T POSSIBLY DO THAT IN SUCH A SHORT TIME I THINK \textbf{<spk2>} DON'T HAVE TO DO THAT \textbf{<spk0>} THAT'S FOR
\vspace{3mm}
\\
DNCASR S1: 
\\
\textbf{<spk1>} YEAH BUT YOU CAN POSSIBLY DO THAT IN SUCH A SHORT TIME I THINK \textbf{<spk0>} DON'T HAVE TO DO THAT
\vspace{3mm}
\\
DNCASR S2+CDA: 
\\
\textbf{<spk1>} YEAH BUT YOU CAN POSSIBLY DO THAT IN SUCH A SHORT TIME I THINK \textbf{<spk2>} DON'T HAVE TO DO THAT
\\
\\
\\
\textbf{Example 5:}
This example shows that ASR predicts all words and speaker turns correctly. DNCASR S1 predicts the correct speaker indices but in the wrong order, while DNCASR S2+CDA predicts the correct speaker indices in the correct order.
\vspace{3mm}
\\
Reference: 
\\
\textbf{<spk2>} IT'S COMING AT THE END \textbf{<spk1>} IT'S HERE FORTY EIGHT \textbf{<spk2>} FORTY EIGHT \textbf{<spk3>} AH FORTY EIGHT \textbf{<spk2>} BUT QUITE A LOT OF PEOPLE HAVE SEEN IT ACTUALLY \textbf{<spk3>} YEAH
\vspace{3mm}
\\
DNCASR S1: 
\\
\textbf{<spk2>} IT'S COMING AT THE END \textbf{<spk1>} IT'S HERE FORTY EIGHT \textbf{<spk3>} FORTY EIGHT \textbf{<spk2>} AH FORTY EIGHT \textbf{<spk3>} BUT QUITE A LOT OF PEOPLE HAVE SEEN IT ACTUALLY \textbf{<spk1>} YEAH
\vspace{3mm}
\\
DNCASR S2+CDA: 
\\
\textbf{<spk2>} IT'S COMING AT THE END \textbf{<spk1>} IT'S HERE FORTY EIGHT \textbf{<spk2>} FORTY EIGHT \textbf{<spk3>} AH FORTY EIGHT \textbf{<spk2>} BUT QUITE A LOT OF PEOPLE HAVE SEEN IT ACTUALLY \textbf{<spk3>} YEAH
\\
\\
\\

\pagebreak

\section{Individual meeting cpWER (speaker-attributed WER) Results}
\label{sec:indiv_cpwer}

\cref{tab:individual_cpwer} lists the cpWERs for all of the meetings in the AMI Dev and Eval sets for the DNCASR (S1), DNCASR (S2), and DNCASR (S2+CDA) from \cref{tab:ami_cpwer}. 

\begin{table}[h!]
\begin{center}
\begin{tabular}{cccc}
\toprule[1.5pt]
Meeting ID & S1 & S2 & S2+CDA \\
\midrule
IB4004 & 34.2 & 32.5 & 31.5 \\
IB4002 & 47.7 & 45.0 & 45.3 \\
TS3004d & 41.1 & 38.8 & 39.2 \\
IB4001 & 35.9 & 32.9 & 32.7 \\
TS3004a & 36.1 & 33.1 & 33.6 \\
IS1008c & 26.1 & 24.7 & 24.7 \\
IB4011 & 33.0 & 30.9 & 30.4 \\
IS1008d & 26.0 & 24.2 & 24.2 \\
ES2011b & 29.9 & 29.5 & 28.9 \\
ES2011a & 40.2 & 46.6 & 40.1 \\
IS1008a & 16.9 & 16.7 & 17.1 \\
IS1008b & 18.5 & 17.0 & 17.1 \\
TS3004b & 33.1 & 31.6 & 29.1 \\
IB4003 & 27.0 & 25.1 & 25.3 \\
TS3004c & 37.8 & 34.4 & 33.0 \\
IB4010 & 36.7 & 35.1 & 34.7 \\
ES2011d & 33.6 & 32.5 & 31.8 \\
ES2011c & 31.9 & 26.4 & 26.6 \\
ES2004c & 28.6 & 25.1 & 23.9 \\
ES2004b & 24.1 & 22.0 & 22.4 \\
EN2002a & 47.0 & 46.6 & 42.8 \\
ES2004a & 44.2 & 35.9 & 37.2 \\
ES2004d & 37.6 & 34.0 & 34.1 \\
TS3003d & 34.0 & 31.3 & 31.3 \\
TS3003a & 35.3 & 33.0 & 33.2 \\
EN2002b & 44.4 & 42.1 & 41.4 \\
EN2002c & 38.2 & 34.9 & 34.5 \\
TS3003c & 18.4 & 19.6 & 19.5 \\
EN2002d & 50.1 & 45.3 & 45.5 \\
IS1009b & 27.4 & 27.7 & 27.2 \\
IS1009d & 39.0 & 32.2 & 30.2 \\
IS1009c & 22.4 & 21.2 & 21.4 \\
IS1009a & 38.1 & 34.3 & 34.3 \\
TS3003b & 16.0 & 16.4 & 16.4 \\
\bottomrule
\end{tabular}
\caption{Individual cpWER results on the AMI Dev and Eval sets for DNCASR (S1), DNCASR (S2), and DNCASR (S2+CDA)}
\label{tab:individual_cpwer}
\end{center}
\end{table}

\section{Replacing wavlm-base-plus with wavlm-large as the ASR module encoder}
\label{app:wavlm-large}

\begin{table}[ht]
\begin{center}
\begin{tabular}{ccccc}
\toprule[1.5pt]
\multirow{2}{*}{Model} & \multirow{2}{*}{cpWER} & cpWER \\
& & (Multi) \\
\midrule
DNCASR (base) & 30.7/31.5 & 42.5/44.1 \\
+ oracle words & 14.8/17.4 & 24.4/28.3 \\
\midrule
DNCASR (large) & 27.6/28.0 & 38.0/39.1 \\
+ oracle words & 14.6/16.4 & 23.8/25.3 \\
\bottomrule
\end{tabular}
\caption{\%cpWER and \%cpWER-Multi on AMI, where \%cpWER-Multi is the \%cpWER on multi-talker segments. Comparison of the best DNCASR setup (S2+CDA) from \autoref{tab:ami_cpwer} using either the wavlm-base-plus (base) or wavlm-large (large) model as the Wav encoder. `+ oracle words' indicates using the oracle words sequence instead of decoded word sequence. }
\label{tab:ami_cpwer_large}
\end{center}
\end{table}

\autoref{tab:ami_cpwer_large} presents the results comparing the best DNCASR setup from \autoref{tab:ami_cpwer} with a modified version that uses wavlm-large instead of wavlm-base-plus as the Wav encoder in the ASR module.
Results show that the cpWER can be reduced by over 10\% after using the larger SSL encoder. 
The relative cpWER reductions using decoded words are 10.1\% and 11.1\% on the AMI Dev and Eval sets, respectively. 
On multi-talker segments, the reductions are quite similar, giving 10.6\% and 11.3\% on Dev and Eval sets respectively. 

When using oracle words to generate ASR hidden features, the relative cpWER reductions with wavlm-large are more modest—1.4\% on Dev and 5.7\% on Eval—both under 10\%, suggesting that the model’s ability to recognise speakers given correct words has not significantly improved.
The reductions in cpWER-Multi using oracle words are more substantial, decreases by 2.5\% and 10.6\%, respectively.

Overall, the results show that improvements are more significant in cpWER-Multi compared to cpWER. This suggests that wavlm-large offers better speaker-related representations in segments with multiple speaker turns, enabling the model to assign relative speaker indices more accurately.

%% file: custom.bbl
\begin{thebibliography}{50}
\providecommand{\natexlab}[1]{#1}

\bibitem[{Anguera et~al.(2007)Anguera, Wooters, and Hernando}]{angueraAcousticBeamforming2007}
Xavier Anguera, Chuck Wooters, and Javier Hernando. 2007.
\newblock Acoustic beamforming for speaker diarization of meetings.
\newblock \emph{IEEE Transactions on Audio, Speech, and Language Processing}, 15(7):2011--2022.

\bibitem[{Carletta et~al.(2006)Carletta, Ashby, Bourban, Flynn, Hain, Kadlec, Karaiskos, Kraaij, Kronenthal, Lathoud, Lincoln, Lisowska, and Reidsma}]{carlettaAmiMeeting2006}
Jean Carletta, Simone Ashby, Sebastien Bourban, Mike Flynn, Thomas Hain, Jaroslav Kadlec, Vasilis Karaiskos, Wessel Kraaij, Melissa Kronenthal, Guillaume Lathoud, Mike Lincoln, Agnes Lisowska, and Mccowan Wilfried Post~Dennis Reidsma. 2006.
\newblock {The Ami meeting corpus}: {{A}} pre-announcement.
\newblock In \emph{In {{Proceedings}} of the {{Second}} International {{Workshop}} on {{Machine Learning}} for {{Multimodal Interaction}}}.

\bibitem[{Chang et~al.(2020)Chang, Zhang, Qian, Roux, and Watanabe}]{changEndtoendMultispeaker2020}
Xuankai Chang, Wangyou Zhang, Yanmin Qian, Jonathan~Le Roux, and Shinji Watanabe. 2020.
\newblock End-to-end multi-speaker speech recognition with transformer.
\newblock In \emph{Proc. {{ICASSP}}}, Barcelona, Spain.

\bibitem[{Chen et~al.(2022)Chen, Wang, Chen, Wu, Liu, Chen, Li, Kanda, Yoshioka, Xiao, Wu, Zhou, Ren, Qian, Qian, Wu, Zeng, Yu, and Wei}]{chenWavLMLargescale2022}
Sanyuan Chen, Chengyi Wang, Zhengyang Chen, Yu~Wu, Shujie Liu, Zhuo Chen, Jinyu Li, Naoyuki Kanda, Takuya Yoshioka, Xiong Xiao, Jian Wu, Long Zhou, Shuo Ren, Yanmin Qian, Yao Qian, Jian Wu, Michael Zeng, Xiangzhan Yu, and Furu Wei. 2022.
\newblock \href {https://doi.org/10.1109/JSTSP.2022.3188113} {{WavLM}: Large-scale self-supervised pre-training for full stack speech processing}.
\newblock \emph{IEEE Journal of Selected Topics in Signal Processing}, 16(6):1505--1518.

\bibitem[{Chen and Gopalakrishnan(1998)}]{chenSpeakerEnvironment1998}
Scott~Shaobing Chen and P~S Gopalakrishnan. 1998.
\newblock Speaker, environment and channel change detection and clustering via the bayesian information criterion.
\newblock In \emph{Proceedings {{DARPA}} Broadcast News Transcription and Understanding Workshop}.

\bibitem[{Chen et~al.(2020)Chen, Yoshioka, Lu, Zhou, Meng, Luo, Wu, Xiao, and Li}]{chenContinuousSpeech2020}
Zhuo Chen, Takuya Yoshioka, Liang Lu, Tianyan Zhou, Zhong Meng, Yi~Luo, Jian Wu, Xiong Xiao, and Jinyu Li. 2020.
\newblock Continuous speech separation: Dataset and analysis.
\newblock In \emph{Proc. {{ICASSP}}}, Barcelona, Spain.

\bibitem[{Chung et~al.(2018)Chung, Nagrani, and Zisserman}]{chungVoxCeleb2Deep2018}
Joon~Son Chung, Arsha Nagrani, and Andrew Zisserman. 2018.
\newblock \href {https://doi.org/10.21437/Interspeech.2018-1929} {{{VoxCeleb2}}: Deep speaker recognition}.
\newblock In \emph{Proc. {Interspeech}}, Hyderabad, India.

\bibitem[{Cornell et~al.(2024)Cornell, Jung, Watanabe, and Squartini}]{cornellOneModel2024}
Samuele Cornell, Jee-weon Jung, Shinji Watanabe, and Stefano Squartini. 2024.
\newblock One model to rule them all? {Towards} end-to-end joint speaker diarization and speech recognition.
\newblock In \emph{Proc. {{ICASSP}}}, Seoul, Korea.

\bibitem[{Cornell et~al.(2023)Cornell, Wiesner, Watanabe, Raj, Chang, Garcia, Maciejewski, Masuyama, Wang, Squartini et~al.}]{cornell2023chime}
Samuele Cornell, Matthew Wiesner, Shinji Watanabe, Desh Raj, Xuankai Chang, Paola Garcia, Matthew Maciejewski, Yoshiki Masuyama, Zhong-Qiu Wang, Stefano Squartini, et~al. 2023.
\newblock The {CHiME}-7 {DASR} challenge: Distant meeting transcription with multiple devices in diverse scenarios.
\newblock In \emph{Proc. {CHiME}}.

\bibitem[{Dawalatabad et~al.(2021)Dawalatabad, Ravanelli, Grondin, Thienpondt, Desplanques, and Na}]{dawalatabadECAPATDNNEmbeddings2021}
Nauman Dawalatabad, Mirco Ravanelli, Fran{\c c}ois Grondin, Jenthe Thienpondt, Brecht Desplanques, and Hwidong Na. 2021.
\newblock \href {https://doi.org/10.21437/Interspeech.2021-941} {{{ECAPA-TDNN}} embeddings for speaker diarization}.
\newblock In \emph{Proc. {{Interspeech}}}, Brno, Czech Republic.

\bibitem[{Dehak et~al.(2011)Dehak, Kenny, Dehak, Dumouchel, and Ouellet}]{dehakFrontendFactor2011}
Najim Dehak, Patrick~J. Kenny, R{\'e}da Dehak, Pierre Dumouchel, and Pierre Ouellet. 2011.
\newblock Front-end factor analysis for speaker verification.
\newblock \emph{IEEE Transactions on Audio, Speech, and Language Processing}, 19(4):788--798.

\bibitem[{Diaconis and Shahshahani(1987)}]{diaconis1987}
Persi Diaconis and Mehrdad Shahshahani. 1987.
\newblock {The subgroup algorithm for generating uniform random variables}.
\newblock \emph{Probability in the Engineering and Informational Sciences}, 1:15--32.

\bibitem[{Fujita et~al.(2019)Fujita, Kanda, Horiguchi, Nagamatsu, and Watanabe}]{Fujita2019Interspeech}
Yusuke Fujita, Naoyuki Kanda, Shota Horiguchi, Kenji Nagamatsu, and Shinji Watanabe. 2019.
\newblock End-to-end neural speaker diarization with permutation-free objectives.
\newblock In \emph{Interspeech}, Brighton, UK.

\bibitem[{Horiguchi et~al.(2022)Horiguchi, Fujita, Watanabe, Xue, and Garcia}]{horiguchiEncoderdecoderBased2022}
Shota Horiguchi, Yusuke Fujita, Shinji Watanabe, Yawen Xue, and Paola Garcia. 2022.
\newblock Encoder-decoder based attractors for end-to-end neural diarization.
\newblock \emph{IEEE/ACM Transactions on Audio, Speech, and Language Processing}, 30:1493--1507.

\bibitem[{Kanda et~al.(2020{\natexlab{a}})Kanda, Gaur, Wang, Meng, Chen, Zhou, and Yoshioka}]{kandaJointSpeaker2020}
Naoyuki Kanda, Yashesh Gaur, Xiaofei Wang, Zhong Meng, Zhuo Chen, Tianyan Zhou, and Takuya Yoshioka. 2020{\natexlab{a}}.
\newblock Joint speaker counting, speech recognition, and speaker identification for overlapped speech of any number of speakers.
\newblock In \emph{Proc. {{Interspeech}}}, Shanghai, China.

\bibitem[{Kanda et~al.(2020{\natexlab{b}})Kanda, Gaur, Wang, Meng, and Yoshioka}]{kandaSerializedOutput2020}
Naoyuki Kanda, Yashesh Gaur, Xiaofei Wang, Zhong Meng, and Takuya Yoshioka. 2020{\natexlab{b}}.
\newblock Serialized output training for end-to-end overlapped speech recognition.
\newblock In \emph{Proc. {{Interspeech}}}, Shanghai, China.

\bibitem[{Kanda et~al.(2022)Kanda, Xiao, Gaur, Wang, Meng, Chen, and Yoshioka}]{kandaTranscribediarizeNeural2022}
Naoyuki Kanda, Xiong Xiao, Yashesh Gaur, Xiaofei Wang, Zhong Meng, Zhuo Chen, and Takuya Yoshioka. 2022.
\newblock Transcribe-to-diarize: {{Neural}} speaker diarization for unlimited number of speakers using end-to-end speaker-attributed {{ASR}}.
\newblock In \emph{Proc. {{ICASSP}}}, Singapore.

\bibitem[{Kinoshita et~al.(2021)Kinoshita, Delcroix, and Tawara}]{eend-vector-clustering}
Keisuke Kinoshita, Marc Delcroix, and Naohiro Tawara. 2021.
\newblock Integrating end-to-end neural and clustering-based diarization: Getting the best of both worlds.
\newblock In \emph{Proc. {ICASSP}}, Toronto, Canada.

\bibitem[{Koluguri et~al.(2022)Koluguri, Park, and Ginsburg}]{koluguriTitanetNeural2022}
Nithin~Rao Koluguri, Taejin Park, and Boris Ginsburg. 2022.
\newblock Titanet: Neural model for speaker representation with 1d depth-wise separable convolutions and global context.
\newblock In \emph{Proc. {{ICASSP}}}, Singapore.

\bibitem[{Landini et~al.(2024)Landini, Diez, Stafylakis, and Burget}]{landiniDiaPerEndEnd2024}
Federico Landini, Mireia Diez, Themos Stafylakis, and Luk{\'a}{\v s} Burget. 2024.
\newblock \href {https://doi.org/10.1109/TASLP.2024.3422818} {{{DiaPer}}: {{End-to-End neural diarization with perceiver-based attractors}}}.
\newblock \emph{IEEE/ACM Transactions on Audio, Speech, and Language Processing}, 32:3450--3465.

\bibitem[{Li et~al.(2021)Li, Kreyssig, Zhang, and Woodland}]{liDiscriminativeNeural2020}
Qiujia Li, Florian~L. Kreyssig, Chao Zhang, and Philip~C. Woodland. 2021.
\newblock Discriminative neural clustering for speaker diarisation.
\newblock In \emph{Proc. {{SLT}}}, Shenzhen, China.

\bibitem[{Lu et~al.(2021{\natexlab{a}})Lu, Kanda, Li, and Gong}]{luStreamingEndtoend2021}
Liang Lu, Naoyuki Kanda, Jinyu Li, and Yifan Gong. 2021{\natexlab{a}}.
\newblock Streaming end-to-end multi-talker speech recognition.
\newblock \emph{IEEE Signal Processing Letters}, 28:803--807.

\bibitem[{Lu et~al.(2021{\natexlab{b}})Lu, Kanda, Li, and Gong}]{luStreamingMultitalker2021}
Liang Lu, Naoyuki Kanda, Jinyu Li, and Yifan Gong. 2021{\natexlab{b}}.
\newblock \href {https://arxiv.org/abs/2104.02109} {Streaming multi-talker speech recognition with joint speaker identification}.
\newblock In \emph{Proc. {{Interspeech}}}, Brno, Czech Republic.

\bibitem[{Nagrani et~al.(2020)Nagrani, Chung, Xie, and Zisserman}]{nagraniVoxcelebLargescale2020}
Arsha Nagrani, Joon~Son Chung, Weidi Xie, and Andrew Zisserman. 2020.
\newblock \href {https://doi.org/10.1016/j.csl.2019.101027} {Voxceleb: Large-scale speaker verification in the wild}.
\newblock \emph{Computer Speech \& Language}, 60:101027.

\bibitem[{Ning et~al.(2006)Ning, Liu, Tang, and Huang}]{ningSpectralClustering2006}
Huazhong Ning, Ming Liu, Hao Tang, and Thomas~S. Huang. 2006.
\newblock A spectral clustering approach to speaker diarization.
\newblock In \emph{Interspeech 2006}, Pittsburgh, USA.

\bibitem[{Panayotov et~al.(2015)Panayotov, Chen, Povey, and Khudanpur}]{panayotovLibrispeechASR2015}
Vassil Panayotov, Guoguo Chen, Daniel Povey, and Sanjeev Khudanpur. 2015.
\newblock \href {https://doi.org/10.1109/ICASSP.2015.7178964} {Librispeech: An {{ASR}} corpus based on public domain audio books}.
\newblock In \emph{Proc. {{ICASSP}}}, Brisbane, Australia.

\bibitem[{Park et~al.(2021)Park, Kanda, Dimitriadis, Han, Watanabe, and Narayanan}]{parkReviewSpeaker2021}
Tae~Jin Park, Naoyuki Kanda, Dimitrios Dimitriadis, Kyu~J. Han, Shinji Watanabe, and Shrikanth Narayanan. 2021.
\newblock \href {https://arxiv.org/abs/2101.09624} {A review of speaker diarization: Recent advances with deep learning}.
\newblock \emph{Computer Speech \& Language}.

\bibitem[{Prabhavalkar et~al.(2024)Prabhavalkar, Hori, Sainath, Schl{\"u}ter, and Watanabe}]{prabhavalkarEndtoendSpeech2024}
Rohit Prabhavalkar, Takaaki Hori, Tara~N. Sainath, Ralf Schl{\"u}ter, and Shinji Watanabe. 2024.
\newblock End-to-end speech recognition: A survey.
\newblock \emph{IEEE/ACM Transactions on Audio, Speech, and Language Processing}, 32:325--351.

\bibitem[{Raj et~al.(2021)Raj, Denisov, Chen, Erdogan, Huang, He, Watanabe, Du, Yoshioka, Luo, Kanda, Li, Wisdom, and Hershey}]{rajIntegrationSpeech2021}
Desh Raj, Pavel Denisov, Zhuo Chen, Hakan Erdogan, Zili Huang, Maokui He, Shinji Watanabe, Jun Du, Takuya Yoshioka, Yi~Luo, Naoyuki Kanda, Jinyu Li, Scott Wisdom, and John~R. Hershey. 2021.
\newblock Integration of speech separation, diarization, and recognition for multi-speaker meetings: System description, comparison, and analysis.
\newblock In \emph{Proc. {{SLT}}}, Shenzhen, China.

\bibitem[{Raj et~al.(2023)Raj, Povey, and Khudanpur}]{rajSURT202023}
Desh Raj, Daniel Povey, and Sanjeev Khudanpur. 2023.
\newblock {{SURT}} 2.0: {{Advances}} in transducer-based multi-talker speech recognition.
\newblock \emph{IEEE/ACM Transactions on Audio, Speech, and Language Processing}, 31:3800--3813.

\bibitem[{Seki et~al.(2018)Seki, Hori, Watanabe, Le~Roux, and Hershey}]{sekiPurelyEndtoend2018}
Hiroshi Seki, Takaaki Hori, Shinji Watanabe, Jonathan Le~Roux, and John~R. Hershey. 2018.
\newblock A purely end-to-end system for multi-speaker speech recognition.
\newblock In \emph{Proc. {{ACL}}}, Melbourne, Australia. Association for Computational Linguistics.

\bibitem[{Sell et~al.(2018)Sell, Snyder, McCree, {Garcia-Romero}, Villalba, Maciejewski, Manohar, Dehak, Povey, Watanabe, and Khudanpur}]{sellDiarizationHard2018}
Gregory Sell, David Snyder, Alan McCree, Daniel {Garcia-Romero}, Jes{\'u}s Villalba, Matthew Maciejewski, Vimal Manohar, Najim Dehak, Daniel Povey, Shinji Watanabe, and Sanjeev Khudanpur. 2018.
\newblock Diarization is hard: {{Some}} experiences and lessons learned for the {{JHU}} team in the inaugural {{DIHARD}} challenge.
\newblock In \emph{Proc. {{Interspeech}}}, Hyderabad, India.

\bibitem[{Shafey et~al.(2019)Shafey, Soltau, and Shafran}]{shafeyJointSpeech2019}
Laurent~El Shafey, Hagen Soltau, and Izhak Shafran. 2019.
\newblock Joint speech recognition and speaker diarization via sequence transduction.
\newblock In \emph{Proc. {{Interspeech}}}, Graz, Austria.

\bibitem[{Sklyar et~al.(2021)Sklyar, Piunova, and Liu}]{sklyarStreamingMultispeaker2021}
Ilya Sklyar, Anna Piunova, and Yulan Liu. 2021.
\newblock Streaming multi-speaker {{ASR}} with {{RNN-T}}.
\newblock In \emph{Proc. {{ICASSP}}}, Toronto, Canada.

\bibitem[{Sklyar et~al.(2022)Sklyar, Piunova, Zheng, and Liu}]{sklyarMultiturnRNNt2022}
Ilya Sklyar, Anna Piunova, Xianrui Zheng, and Yulan Liu. 2022.
\newblock Multi-turn {{RNN-t}} for streaming recognition of multi-party speech.
\newblock In \emph{Proc. {{ICASSP}}}, Singapore.

\bibitem[{Snyder et~al.(2018)Snyder, {Garcia-Romero}, Sell, Povey, and Khudanpur}]{snyderXvectorsRobust2018}
David Snyder, Daniel {Garcia-Romero}, Gregory Sell, Daniel Povey, and Sanjeev Khudanpur. 2018.
\newblock X-vectors: Robust {{DNN}} embeddings for speaker recognition.
\newblock In \emph{Proc. {{ICASSP}}}, Calgary, Canada.

\bibitem[{Sohn et~al.(1999)Sohn, Kim, and Sung}]{sohnStatisticalModelbased1999}
Jongseo Sohn, Nam~Soo Kim, and Wonyong Sung. 1999.
\newblock A statistical model-based voice activity detection.
\newblock \emph{IEEE Signal Processing Letters}, 6(1):1--3.

\bibitem[{Stewart(1980)}]{stewartEfficientGenerationRandom1980}
G.~W. Stewart. 1980.
\newblock The efficient generation of random orthogonal matrices with an application to condition estimators.
\newblock \emph{SIAM Journal on Numerical Analysis}, 17(3).

\bibitem[{Sun et~al.(2021)Sun, Zhang, and Woodland}]{sunCombinationDeep2021}
Guangzhi Sun, Chao Zhang, and Phil Woodland. 2021.
\newblock \href {https://doi.org/10.1016/j.neunet.2021.04.020} {Combination of deep speaker embeddings for diarisation}.
\newblock \emph{Neural Networks}, 141:372--384.

\bibitem[{Tranter and Reynolds(2006)}]{tranterOverviewAutomatic2006}
S.E. Tranter and D.A. Reynolds. 2006.
\newblock An overview of automatic speaker diarization systems.
\newblock \emph{IEEE Transactions on Audio, Speech, and Language Processing}, 14(5):1557--1565.

\bibitem[{Vaswani et~al.(2017)Vaswani, Shazeer, Parmar, Uszkoreit, Jones, Gomez, Kaiser, and Polosukhin}]{vaswaniAttentionAll2017}
Ashish Vaswani, Noam Shazeer, Niki Parmar, Jakob Uszkoreit, Llion Jones, Aidan~N Gomez, {\L}ukasz Kaiser, and Illia Polosukhin. 2017.
\newblock Attention is all you need.
\newblock In \emph{Proc. {{NIPS}}}, Long Beach, USA.

\bibitem[{{von Neumann} et~al.(2023){von Neumann}, Boeddeker, Delcroix, and {Haeb-Umbach}}]{vonneumannMeetEvalToolkit2023}
Thilo {von Neumann}, Christoph Boeddeker, Marc Delcroix, and Reinhold {Haeb-Umbach}. 2023.
\newblock \href {https://doi.org/10.48550/arXiv.2307.11394} {{{MeetEval}}: {{A toolkit}} for computation of word error rates for meeting transcription systems}.
\newblock In \emph{Proc. {{CHiME}}}.

\bibitem[{Wang et~al.(2016)Wang, Zhang, Woodland, Gales, Karanasou, Lanchantin, Liu, and Qian}]{wangImprovedDNNbased2016}
L.~Wang, C.~Zhang, P.~C. Woodland, M.~J.~F. Gales, P.~Karanasou, P.~Lanchantin, X.~Liu, and Y.~Qian. 2016.
\newblock Improved {{DNN-based}} segmentation for multi-genre broadcast audio.
\newblock In \emph{Proc. {{ICASSP}}}, Shanghai, China.

\bibitem[{Watanabe et~al.(2017)Watanabe, Hori, Kim, Hershey, and Hayashi}]{watanabeHybridCTC2017}
Shinji Watanabe, Takaaki Hori, Suyoun Kim, John~R. Hershey, and Tomoki Hayashi. 2017.
\newblock Hybrid {{CTC}}/attention architecture for end-to-end speech recognition.
\newblock \emph{IEEE Journal of Selected Topics in Signal Processing}, 11(8):1240--1253.

\bibitem[{Wilcoxon(1945)}]{wilcoxonIndividualComparisons1945}
Frank Wilcoxon. 1945.
\newblock \href {https://arxiv.org/abs/3001968} {Individual comparisons by ranking methods}.
\newblock \emph{Biometrics Bulletin}, 1(6):80--83.

\bibitem[{Wolf et~al.(2020)Wolf, Debut, Sanh, Chaumond, Delangue, Moi, Cistac, Rault, Louf, Funtowicz, Davison, Shleifer, {von Platen}, Ma, Jernite, Plu, Xu, Scao, Gugger, Drame, Lhoest, and Rush}]{wolfTransformersStateoftheart2020}
Thomas Wolf, Lysandre Debut, Victor Sanh, Julien Chaumond, Clement Delangue, Anthony Moi, Pierric Cistac, Tim Rault, R{\'e}mi Louf, Morgan Funtowicz, Joe Davison, Sam Shleifer, Patrick {von Platen}, Clara Ma, Yacine Jernite, Julien Plu, Canwen Xu, Teven~Le Scao, Sylvain Gugger, Mariama Drame, Quentin Lhoest, and Alexander~M. Rush. 2020.
\newblock Transformers: {{State-of-the-art}} natural language processing.
\newblock In \emph{Proc. {{EMNLP}}}.

\bibitem[{Young et~al.(2015)Young, Evermann, Gales, Hain, Kershaw, Liu, Moore, Odell, Ollason, Povey, Ragni, Valtchev, Woodland, and Zhang}]{youngHTKBook2015}
Steve Young, Gunnar Evermann, Mark Gales, Thomas Hain, Dan Kershaw, Xunying~(Andrew) Liu, Gareth Moore, Julian Odell, Dave Ollason, Dan Povey, Anton Ragni, Valtcho Valtchev, Phil Woodland, and Chao Zhang. 2015.
\newblock The {{HTK}} book.
\newblock \emph{University of Cambridge}.

\bibitem[{Zhang et~al.(2019)Zhang, Wang, Zhu, Paisley, and Wang}]{zhangFullySupervised2019}
Aonan Zhang, Quan Wang, Zhenyao Zhu, John Paisley, and Chong Wang. 2019.
\newblock Fully supervised speaker diarization.
\newblock In \emph{Proc. {{ICASSP}}}, Brighton, UK.

\bibitem[{Zheng et~al.(2024)Zheng, Sun, Zhang, and Woodland}]{zhengSOTTriggered2024}
Xianrui Zheng, Guangzhi Sun, Chao Zhang, and Philip~C. Woodland. 2024.
\newblock {{SOT}} triggered neural clustering for speaker attributed {{ASR}}.
\newblock In \emph{Proc. {{Interspeech}}}, Kos Island, Greece.

\bibitem[{Zheng et~al.(2022)Zheng, Zhang, and Woodland}]{zhengTandemMultitask2022}
Xianrui Zheng, Chao Zhang, and Philip~C. Woodland. 2022.
\newblock Tandem multitask training of speaker diarisation and speech recognition for meeting transcription.
\newblock In \emph{Proc. {{Interspeech}}}, Incheon, Korea.

\end{thebibliography}
